\documentclass[conference]{IEEEtran}
\IEEEoverridecommandlockouts
\usepackage{cite}
\usepackage{amsmath,amssymb,amsfonts}
\usepackage{algorithmic}
\usepackage{graphicx}
\usepackage{textcomp}
\usepackage{xcolor}

\usepackage{booktabs}
\usepackage{caption}
\usepackage{array}
\usepackage{float}

\def\BibTeX{{\rm B\kern-.05em{\sc i\kern-.025em b}\kern-.08em
    T\kern-.1667em\lower.7ex\hbox{E}\kern-.125emX}}
\begin{document}

\title{SACNet: A Spatially Adaptive Convolution Network for 2D
	Multi-organ Medical  Segmentation\\
\thanks{*Corresponding author.This work is supported by National Natural Science
Foundation of China [No. 62371156] and Natural Science Foundation of Guangdong Province [No. 2022A1515011629]}

}

\author{\IEEEauthorblockN{Lin Zhang, Wenbo Gao, Jie Yi, Yunyun Yang$^{*}$}
\IEEEauthorblockA{\textit{School of Science} \\
\textit{Harbin Institute of Technology, Shenzhen}\\
Shenzhen, China \\
23s058005@stu.hit.edu.cn, 23B358001@stu.hit.edu.cn, tomatosquid@163.com, yangyunyun@hit.edu.cn}
}

\maketitle

\begin{abstract}
Multi-organ segmentation in medical image analysis is crucial for diagnosis and treatment planning. However, many factors complicate the task, including variability in different target categories and interference from complex backgrounds. In this paper, we utilize the knowledge of Deformable Convolution V3 (DCNv3) and multi-object segmentation to optimize our Spatially Adaptive Convolution Network (SACNet) in three aspects: feature extraction, model architecture, and loss constraint, simultaneously enhancing the perception of different segmentation targets. Firstly, we propose the Adaptive Receptive Field Module (ARFM), which combines DCNv3 with a series of customized block-level and architecture-level designs similar to transformers. This module can capture the unique features of different organs by adaptively adjusting the receptive field according to various targets. Secondly, we utilize ARFM as building blocks to construct the encoder-decoder of SACNet and partially share parameters between the encoder and decoder, making the network wider rather than deeper. This design achieves a shared lightweight decoder and a more parameter-efficient and effective framework. Lastly, we propose a novel continuity dynamic adjustment loss function, based on t-vMF dice loss and cross-entropy loss, to better balance easy and complex classes in segmentation. Experiments on 3D slice datasets from Synapse demonstrate that SACNet delivers superior segmentation performance in multi-organ segmentation tasks compared to several existing methods.
\end{abstract}


\section{Introduction}

Multi-organ medical image segmentation, which simultaneously provides clearer visualization of anatomical and pathological structures across multiple organs, significantly enhances diagnostic efficiency and accuracy \cite{b1}-\cite{b2}. 

However, it remains challenging due to several difficulties: (1) Variability across target categories. Not only do different organs within the same dataset vary significantly in size, shape, and texture, but even different slices of the same organ can exhibit considerable differences. This variability makes it challenging for segmentation models to generalize effectively, as they may overfit to specific features seen during training and struggle with unseen morphological variations. (2) Interference from complex backgrounds. The complexity of surrounding anatomical structures and the presence of varying textures, intensities, and overlapping tissues introduce substantial challenges. These intricate backgrounds can lead to ambiguous boundaries and obscure the precise contours of target organs, compromising the model's ability to accurately segment organs and often resulting in incomplete or inaccurate segmentation outcomes.

To address these challenges in multi-organ medical image segmentation, recent studies \cite{b3}-\cite{b16} have employed various methods, including standard, dilated, and large-kernel convolutions, transformers, and hybrid architectures, to improve segmentation accuracy. Dilated and large-kernel convolutions expand the receptive field, while transformers \cite{b9} utilize self-attention to capture long-range dependencies. However, these models often lack adaptability to the specific structures of different organs, leading to potential segmentation inaccuracies. Some researchers \cite{b17}-\cite{b20} have incorporated deformable convolutions and deformable attention transformers to address these issues, though challenges such as sampling flexibility, computational complexity, and data requirements remain. Additionally, enhancements to the Dice loss function \cite{b21}-\cite{b26} have been proposed to handle segmentation imbalances, but limitations persist.

We propose the SACNet framework, which features an Adaptive Receptive Field Module (ARFM), a widenet strategy, and a dynamic continuity adjustment loss function:

\begin{itemize}
    \item \textbf{ARFM}: To handle varying target categories and complex backgrounds, we combine DCNv3 with grouped convolutions, ensuring each segmentation target has an adaptive receptive field. Additional elements like Feed-Forward, Layer Scale, and DropPath blocks enhance stability and geometric perception.
    \item \textbf{WideNet Strategy}: To reduce computational load, we share DCNv3 block weights between encoder and decoder, allowing the network to be wider instead of deeper, achieving a lightweight and efficient design.
    \item \textbf{Continuity Dynamic Adjustment Loss Function (CTLoss)}: Combining t-vMF Dice loss and cross-entropy loss, CTLoss adjusts based on the Intersection Over Union (IOU) after each epoch, handling imbalanced data by dynamically adjusting the similarities for different classes.
\end{itemize}

Our method, SACNet, outperforms existing approaches in multi-organ segmentation tasks, as demonstrated by state-of-the-art results and thorough ablation studies.

\begin{figure}[htbp]
\centering
\includegraphics[width=\columnwidth]{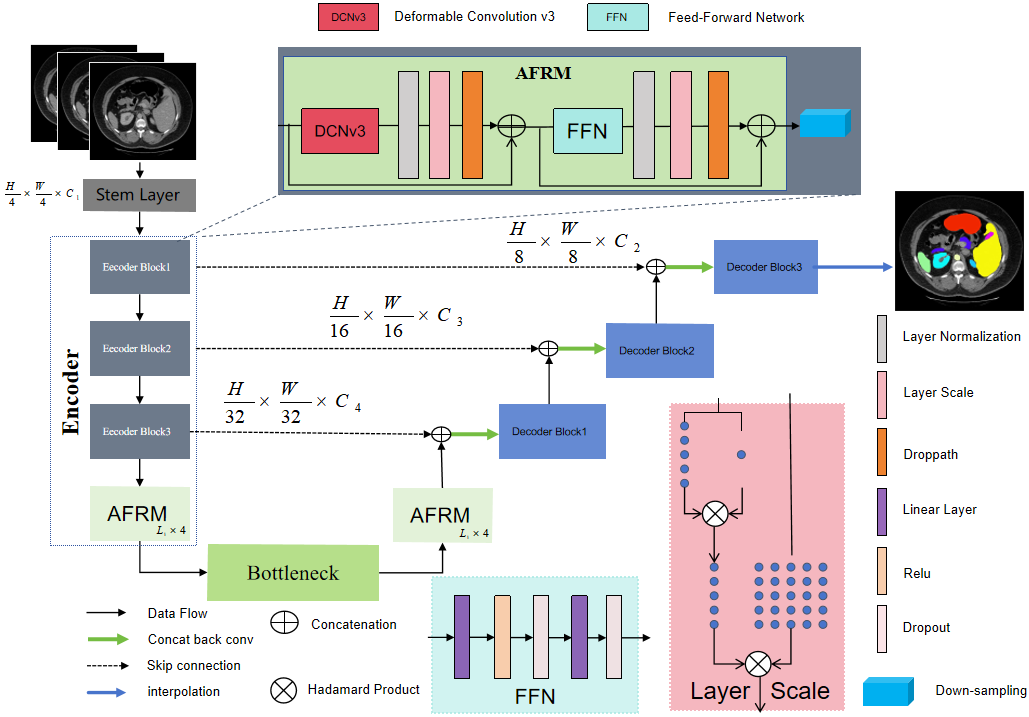}
\caption{Overview of the SACNet architecture.}
\label{fig1}
\end{figure}

\section{METHOD}
\subsection{Encoder}\label{AA}
Recent developments in large-scale CNN models \cite{b27} highlight the potential of Deformable Convolution v3 (DCNv3) for downstream visual tasks. In SACNet, the ARFM module leverages the pre-trained DCNv3 block as a core component, integrating it with Transformer-like designs to form the spatially adaptive pyramid feature encoder. The input image, sized 3×224×224, begins with a Stem Layer for initial processing. The encoder consists of four stages with embedding sizes of 112, 224, 448, and 896, each containing a series of ARFM modules and down-sampling layers. Each ARFM module includes five blocks: DCNv3 for adaptive receptive fields, Layer Norm, Feed-Forward block for pattern extraction, and Layer Scale and DropPath to stabilize training. 

We emphasize the third encoder stage, using an ARFM module ratio of [4, 4, 21, 4] for optimal computation distribution. Skip connections are employed to maintain gradient flow. The encoder produces four feature maps of size \( \{ 1 \times (112 \times 2^i) \times \frac{H}{2^{2+i}} \times \frac{W}{2^{2+i}} \}\), where \( i = \{0, 1, 2, 3\} \), which are linked to the decoder via skip connections. The module sequence is illustrated in Fig.\ref{fig1}.

\subsection{Decoder}
The SACNet decoder is composed of four decoder blocks and one bottleneck block. Each block starts with layer normalization and upsamples the input by a factor of two to match the encoder outputs. The ARFM module refines features, followed by layer normalization, maintaining the same dimensions. The outputs are concatenated with the encoder outputs via skip connections. To handle the growing complexity due to DCNv3's sampling points, we apply the WideNet \cite{b31} strategy, sharing DCNv3 projection weights between encoder and decoder to reduce parameter load. This allows SACNet to expand in width while limiting depth. After upsampling, images are resized to \( H \times W \times N \) (where \( N \) is the number of classes) for final segmentation, followed by Softmax for pixel classification. The bottleneck block functions similarly to other decoder blocks but focuses more on channel information without altering image size. No additional techniques like residual connections \cite{b29} or deep supervision \cite{b30} are used in the decoder.

\begin{table*}[t]
\captionsetup{justification=raggedright,singlelinecheck=false}
\caption{Comparative analysis of model performance with 2D methods in multi-organ segmentation using the synapse dataset. The best result within each column is highlighted in bold, and the second-best is underlined. Models marked with an asterisk ($^{*}$) were implemented by us.}

\begin{center}
\begin{tabular}{lccccccccccccc}
\toprule
Model & Year & Aorta & Gallbladder & Kidney(L) & Kidney(R) & Liver & Pancreas & Spleen & Stomach & \multicolumn{2}{c}{Average} \\
\cmidrule(lr){11-12}
& & & & & & & & & & DSC$\uparrow$ & HD95$\downarrow$ \\
\midrule
U-net \cite{b3} & 2015 & \textbf{89.07} & 69.72 & 77.77 & 68.6 & 93.43 & 53.98 & 86.67 & 75.58 & 76.85\% & 39.70 \\
U-Net++ \cite{b4} & 2018 & 88.19 & 68.89 & 81.76 & 75.27 & 93.01 & 58.20 & 83.44 & 70.52 & 76.91\% & 36.93 \\
MultiResUNet \cite{b5} & 2020 & 87.33 & 65.67 & 82.08 & 73.82 & 93.68 & 52.85 & 85.23 & 75.66 & 77.42\% & 36.84 \\
TransUNet \cite{b9} & 2021 & 87.23 & 63.13 & 81.87 & 77.02 & 94.08 & 55.86 & 85.08 & 75.62 & 77.48\% & 31.69 \\
Swin-UNet \cite{b13} & 2022 & 85.47 & 66.53 & 83.28 & 79.61 & 94.29 & 56.58 & 90.66 & 76.60 & 79.13\% & 21.55 \\
DA-TransUNet$^{*}$ \cite{b11} & 2023 & 87.13 & 58.75 & 79.39 & 78.41 & 94.82 & 62.62 & 87.84 & 79.72 & 78.59\% & 29.20 \\
ParaTransCNN \cite{b12} & 2024 & 88.12 & 68.97 & \underline{87.99} & \underline{83.84} & 95.01 & 69.79 & \textbf{92.71} & 84.43 & \underline{83.86}\% & \underline{15.86} \\
AgileFormer$^{*}$ \cite{b20} & 2024 & 88.52 & 70.37 & 85.69 & 81.14 & 95.65 & \textbf{70.78} & 91.19 & 86.17 & 83.69\% & 19.37 \\
SACNet+\(L_{CD}\)(Ours) & 2024 & 88.46 & 71.05 & 83.68 & 80.99 & \underline{95.74} & \textbf{71.40} & 91.27 & \underline{86.56} & 83.64\% & 22.01 \\
SACNet+\(L_{CT}\)(Ours) & 2024 & \underline{88.88} & \underline{70.45} & \textbf{89.36} & \textbf{85.41} & \textbf{95.98} & \underline{70.38} & 91.26 & \textbf{87.64} & \textbf{84.92\%} & \textbf{15.13} \\
\bottomrule
\end{tabular}
\end{center}
\label{tab1}
\end{table*}

\begin{table*}[t]
\captionsetup{justification=raggedright,singlelinecheck=false}
\caption{Comparison of Ablation Studies: Only a single parameter is altered while keeping the others constant to observe the impact of each parameter on performance. S represents the Layer Scale and DropPath block.}
\begin{center}
\begin{tabular}{lcccccccccccccc}
\toprule
DCNv3 & FFN & S & CFLoss & Aorta & Gallbladder & Kidney(L) & Kidney(R) & Liver & Pancreas & Spleen & Stomach & DSC$\uparrow$ & HD95$\downarrow$ \\
\midrule
           & \checkmark & \checkmark & \checkmark & 84.33 & 60.83 & 85.73 & 77.54 & 93.54 & 50.12 & 88.20 & 75.62 & 76.99\% & 31.05 \\
\checkmark & \checkmark & & \checkmark & 85.81 & 66.21 & 85.58 & 81.09 & 94.98 & 58.81 & 92.16 & 77.12 & 80.22\% & 16.90 \\
\checkmark & & \checkmark & \checkmark & 87.23 & 64.81 & 86.97 & 83.35 & 94.74 & 62.50 & 93.25 & 81.58 & 81.81\% & 15.84 \\
\checkmark & \checkmark & \checkmark & & 88.46 & 71.05 & 83.68 & 80.99 & 95.74 & 71.40 & 91.27 & 86.56 & 83.64\% & 22.01 \\  
\checkmark & \checkmark & \checkmark & \checkmark & 88.88 & 70.45 & 89.36 & 85.41 & 95.98 & 70.38 & 91.26 & 87.64 & 84.92\% & 15.13 \\
\bottomrule
\end{tabular}
\end{center}
\label{tab3}
\end{table*}

\begin{figure}[t]
\centering
\includegraphics[width=\columnwidth]{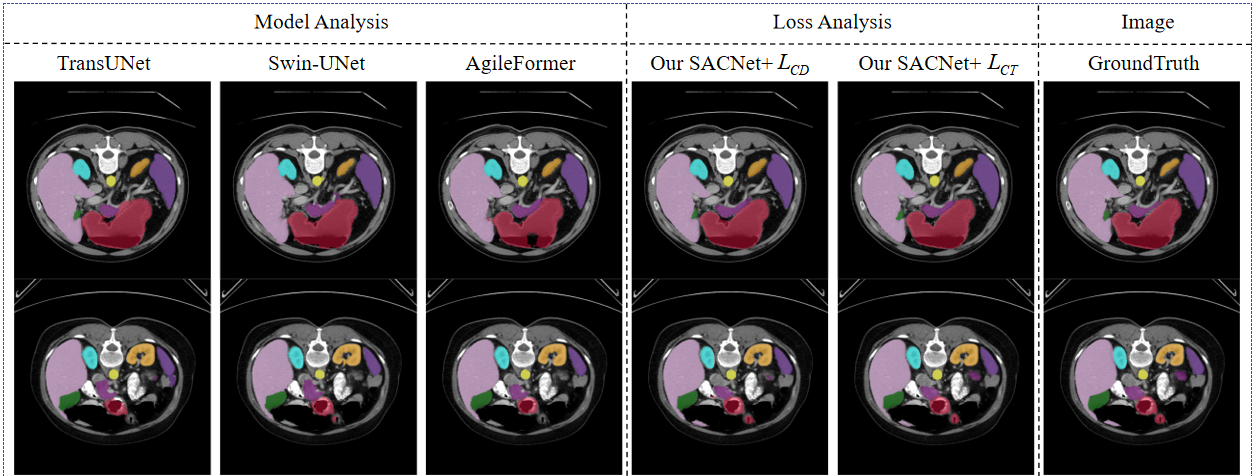}
\caption{Qualitative comparisons of the segmentation performance of our approach alongside other segmentation methods. The first two rows display Synapse dataset results with color codes: yellow (aorta), green (gallbladder), orange (left kidney), cyan (right kidney), plum (liver), purple (pancreas), indigo (spleen), and crimson (stomach). 
}
\label{fig3}
\end{figure}

\subsection{Training Loss}
In this study, we tested most of the proposed loss functions for multi-class medical image segmentation and comprehensively considered the loss aggregation strategy. We propose a novel Continuity Dynamic Adjustment Loss Function (CTLoss) based on t-vMF Dice loss \cite{b25} and the cross-entropy loss function. This new loss function adaptively uses more compact similarities for easy classes and wider similarities for difficult classes, making it more effective in helping the model converge better and more efficiently for multi-object image segmentation.

The t-vMF Dice Loss is formulated as:
\begin{small}
\begin{equation}
\cos\theta_i = \sum_{i=1}^n A_{in} B_{in},
\label{eq:6}
\end{equation}
\begin{equation}
\phi_\kappa(\cos \theta; \kappa) = \frac{1 + \cos \theta}{1 + \kappa (1 - \cos \theta)} - 1,
\label{eq:7}
\end{equation}
\begin{equation}
L_{t-vMF} = \frac{1}{C} \sum_{i=1}^C \left( 1 - \phi_\kappa(\cos \theta_i; \kappa) \right)^2,
\label{eq:8}
\end{equation}
\end{small}
where \( n \) indicates the number of class samples, \( i \) represents the category index, \( A_{in} \) indicates the vectors containing all positive examples predicted by the model, and \( B_{in} \) indicates the vectors containing all positive examples of the ground truth in the dataset. \( \kappa \) is a concentration parameter that adjusts the shape of the similarity function, and \( C \) is the number of classes. We propose an effective algorithm that uses the Intersection Over Union (IOU) of the model on the validation set to update \( \kappa \) in the t-vMF Dice loss after each epoch. This approach adaptively determines \( \kappa \) to achieve more compact similarities for easy classes and wider similarities for difficult classes.

The final loss function combines the adaptive t-vMF Dice Loss and the cross-entropy loss:
\begin{equation}
L_{CT} = \gamma L_{t-vMF} + (1 - \gamma) L_{CE},
\label{eq:10}
\end{equation}
where \( L_{CE} \) is the cross-entropy loss function, \(\gamma\) is the weight coefficient, with \(\gamma = 0.6\) and \(1 - \gamma = 0.4\) representing the weights for the adaptive t-vMF Dice loss and the cross-entropy loss, respectively.

\section{Experiment}
\subsection{Dataset and Evaluation}
The Synapse dataset consists of 30 scans of eight abdominal organs from the MICCAI 2015 Multi-Atlas Abdominal Labeling Challenge. These eight organs include the left kidney, right kidney, aorta, spleen, gallbladder, liver, stomach, and pancreas. We report the mean Dice Similarity Coefficient (DSC) and the 95\% Hausdorff Distance (HD95) for the segmentation of abdominal organs using 18 training cases (2212 axial slices) and 12 validation cases.

\subsection{Setups}
The model and experiments were implemented using PyTorch 2.2.0 and the CUDA toolkit 12.1, with all experiments conducted on an NVIDIA 3090 24G GPU. To improve computational efficiency, the resolution of all training, validation, and test images was resized to 224×224. We used a batch size of 18, an initial learning rate of $3 \times 10^{-4}$, and trained the models for 400 epochs.For better model initialization, we initialized the model parameters with weights pre-trained on ImageNet-1k.

\section{Result}
\subsection{Performance on Synapse Dataset}
On the Synapse dataset, as shown in Table \ref{tab1}, our SACNet stands out with the highest scores for most organs, achieving 89.36\% for the left kidney, 85.41\% for the right kidney, 95.98\% for the liver, and 87.64\% for the stomach. For other organs, except the spleen, our SACNet also achieves the second-best scores. Additionally, our DSC reaches 84.92\% and HD95 is 15.13, both being the best results compared to other benchmarks. Figure \ref{fig3} provides visual examples of the segmentation outputs produced by our proposed model and other state-of-the-art methods. As shown, our predictions have clearer boundaries between different segmentation targets: the boundaries between the pancreas and spleen are clear and smooth. Our model is more accurate for small objects: Swin-UNet and AgileFormer miss the gallbladder in their segmentations, and TransUNet only partially segments it, but our model segments it completely. As shown in the last two rows of Table \ref{tab1}, replacing CDLoss with CTLoss significantly improves the results for DSC and HD95, as well as for most organs, particularly the left and right kidneys.

\subsection{Ablation Study}
Using the Synapse dataset, we ablated SACNet’s key components to validate their effectiveness, as detailed in Table \ref{tab3}. We evaluated the impact of removing pre-trained DCNv3, FFN, Layer Scale, and DropPath blocks. Without DCNv3, DSC dropped from 84.92\% to 76.99\%, and HD95 rose from 15.13 to 31.05, indicating its crucial role. We also compared two loss functions: a combination of cross-entropy and Dice loss versus our proposed CTLoss, with the latter showing superior results. These comparisons confirm that our ARFM design, integrating key components, enhances accuracy, positioning SACNet as a state-of-the-art model for 2D multi-organ segmentation.

\section{Conclusion}
In this study, we focus on the 2D multi-organ medical segmentation task and utilize the knowledge of deformable convolution v3 and multi-object segmentation to optimize our SACNet in three aspects: feature extraction, model architecture, and loss constraint, simultaneously enhancing perception for different segmentation targets. Our method is verified on 3D slice datasets from Synapse, and the results show that our method provides better accuracy in multi-organ segmentation tasks compared to several existing methods.


\begin{thebibliography}{00}

\bibitem{b1} R. Wang, T. Lei, R. Cui, et al., "Medical image segmentation using deep learning: A survey," \textit{IET Image Processing}, vol. 16, no. 5, pp. 1243-1267, 2022.
\bibitem{b2} X. Liu, L. Song, S. Liu, et al., "A review of deep-learning-based medical image segmentation methods," \textit{Sustainability}, vol. 13, no. 3, p. 1224, 2021.
\bibitem{b3} O. Ronneberger, P. Fischer, and T. Brox, "U-net: Convolutional networks for biomedical image segmentation," in \textit{Proc. Medical Image Computing and Computer-Assisted Intervention (MICCAI)}, Munich, Germany, 2015, pp. 234-241.
\bibitem{b4} Z. Zhou, M. M. R. Siddiquee, N. Tajbakhsh, et al., "Unet++: A nested u-net architecture for medical image segmentation," in \textit{Proc. Deep Learning in Medical Image Analysis and Multimodal Learning for Clinical Decision Support (DLMIA)}, Granada, Spain, 2018, pp. 3-11.
\bibitem{b5} N. Ibtehaz and M. S. Rahman, "MultiResUNet: Rethinking the U-Net architecture for multimodal biomedical image segmentation," \textit{Neural Networks}, vol. 121, pp. 74-87, 2020.
\bibitem{b6} Y. Chen, T. Zhou, Y. Chen, et al., "HADCNet: Automatic segmentation of COVID-19 infection based on a hybrid attention dense connected network with dilated convolution," \textit{Computers in Biology and Medicine}, vol. 149, p. 105981, 2022.
\bibitem{b7} X. Shu, Y. Gu, X. Zhang, et al., "FCRB U-Net: A novel fully connected residual block U-Net for fetal cerebellum ultrasound image segmentation," \textit{Computers in Biology and Medicine}, vol. 148, p. 105693, 2022.
\bibitem{b8} F. Isensee, P. F. Jaeger, S. A. A. Kohl, et al., "nnU-Net: a self-configuring method for deep learning-based biomedical image segmentation," \textit{Nature Methods}, vol. 18, no. 2, pp. 203-211, 2021.
\bibitem{b9} J. Chen, Y. Lu, Q. Yu, et al., "Transunet: Transformers make strong encoders for medical image segmentation," \textit{arXiv preprint arXiv:2102.04306}, 2021.
\bibitem{b10} R. Azad, M. T. Al-Antary, M. Heidari, et al., "Transnorm: Transformer provides a strong spatial normalization mechanism for a deep segmentation model," \textit{IEEE Access}, vol. 10, pp. 108205-108215, 2022.
\bibitem{b11} G. Sun, Y. Pan, W. Kong, et al., "DA-TransUNet: Integrating Spatial and Channel Dual Attention with Transformer U-Net for Medical Image Segmentation," \textit{arXiv preprint arXiv:2310.12570}, 2023.
\bibitem{b12} H. Sun, J. Xu, and Y. Duan, "ParaTransCNN: Parallelized TransCNN Encoder for Medical Image Segmentation," \textit{arXiv preprint arXiv:2401.15307}, 2024.
\bibitem{b13} H. Cao, Y. Wang, J. Chen, et al., "Swin-unet: Unet-like pure transformer for medical image segmentation," in \textit{Proc. European Conference on Computer Vision (ECCV)}, Cham, Switzerland: Springer Nature Switzerland, 2022, pp. 205-218.
\bibitem{b14} X. Huang, Z. Deng, D. Li, et al., "MISSFormer: an effective transformer for 2D medical image segmentation," \textit{IEEE Transactions on Medical Imaging}, 2022.
\bibitem{b15} R. Azad, M. Heidari, M. Shariatnia, et al., "Transdeeplab: Convolution-free transformer-based deeplab v3+ for medical image segmentation," in \textit{Proc. International Workshop on PRedictive Intelligence In MEdicine (PRIME)}, Cham, Switzerland: Springer Nature Switzerland, 2022, pp. 91-102.
\bibitem{b16} R. Azad, R. Arimond, E. K. Aghdam, et al., "Dae-former: Dual attention-guided efficient transformer for medical image segmentation," in \textit{Proc. International Workshop on PRedictive Intelligence In MEdicine (PRIME)}, Cham, Switzerland: Springer Nature Switzerland, 2023, pp. 83-95.
\bibitem{b17} J. Dai, H. Qi, Y. Xiong, et al., "Deformable convolutional networks," in \textit{Proc. IEEE International Conference on Computer Vision (ICCV)}, 2017, pp. 764-773.
\bibitem{b18} H. H. Lee, Q. Liu, Q. Yang, et al., "DeformUX-Net: Exploring a 3D Foundation Backbone for Medical Image Segmentation with Depthwise Deformable Convolution," \textit{arXiv preprint arXiv:2310.00199}, 2023.
\bibitem{b19} Z. Xia, X. Pan, S. Song, et al., "Vision transformer with deformable attention," in \textit{Proc. IEEE/CVF Conference on Computer Vision and Pattern Recognition (CVPR)}, 2022, pp. 4794-4803.
\bibitem{b20} P. Qiu, J. Yang, S. Kumar, et al., "AgileFormer: Spatially Agile Transformer UNet for Medical Image Segmentation," \textit{arXiv preprint arXiv:2404.00122}, 2024.
\bibitem{b21} C. H. Sudre, W. Li, T. Vercauteren, et al., "Generalised dice overlap as a deep learning loss function for highly unbalanced segmentations," in \textit{Proc. Deep Learning in Medical Image Analysis and Multimodal Learning for Clinical Decision Support (DLMIA)}, Québec City, QC, Canada, 2017, pp. 240-248.
\bibitem{b22} S. Shit, J. C. Paetzold, A. Sekuboyina, et al., "clDice-a novel topology-preserving loss function for tubular structure segmentation," in \textit{Proc. IEEE/CVF Conference on Computer Vision and Pattern Recognition (CVPR)}, 2021, pp. 16560-16569.
\bibitem{b23} L. Wang, C. Wang, Z. Sun, et al., "An improved dice loss for pneumothorax segmentation by mining the information of negative areas," \textit{IEEE Access}, vol. 8, pp. 167939-167949, 2020.
\bibitem{b24} P. Wang and A. C. S. Chung, "Focal dice loss and image dilation for brain tumor segmentation," in \textit{Proc. Deep Learning in Medical Image Analysis and Multimodal Learning for Clinical Decision Support (DLMIA)}, Granada, Spain, 2018, pp. 119-127.
\bibitem{b25} S. Kato and K. Hotta, "Adaptive t-vmf dice loss for multi-class medical image segmentation," \textit{arXiv preprint arXiv:2207.07842}, 2022.
\bibitem{b26} X. Zhu, H. Hu, S. Lin, et al., "Deformable convnets v2: More deformable, better results," in \textit{Proc. IEEE/CVF Conference on Computer Vision and Pattern Recognition (CVPR)}, 2019, pp. 9308-9316.
\bibitem{b27} W. Wang, J. Dai, Z. Chen, et al., "Internimage: Exploring large-scale vision foundation models with deformable convolutions," in \textit{Proc. IEEE/CVF Conference on Computer Vision and Pattern Recognition (CVPR)}, 2023, pp. 14408-14419.
\bibitem{b28} H. Wang, P. Cao, J. Wang, et al., "Uctransnet: rethinking the skip connections in u-net from a channel-wise perspective with transformer," in \textit{Proc. AAAI Conference on Artificial Intelligence (AAAI)}, 2022, vol. 36, no. 3, pp. 2441-2449.
\bibitem{b29} A. Hatamizadeh, Y. Tang, V. Nath, et al., "Unetr: Transformers for 3d medical image segmentation," in \textit{Proc. IEEE/CVF Winter Conference on Applications of Computer Vision (WACV)}, 2022, pp. 574-584.
\bibitem{b30} B. Dong, W. Wang, D. P. Fan, et al., "Polyp-pvt: Polyp segmentation with pyramid vision transformers," \textit{arXiv preprint arXiv:2108.06932}, 2021.
\bibitem{b31} F. Xue, Z. Shi, F. Wei, et al., "Go wider instead of deeper," in \textit{Proc. AAAI Conference on Artificial Intelligence (AAAI)}, 2022, vol. 36, no. 8, pp. 8779-8787.
\bibitem{b32} H. Wang, S. Xie, L. Lin, et al., "Mixed Transformer U-Net for Medical Image Segmentation," in \textit{Proc. IEEE/CVF Conference on Computer Vision and Pattern Recognition (CVPR)}, 2022, pp. 2390-2394.
\bibitem{b33} M. M. Rahman and R. Marculescu, "Medical image segmentation via cascaded attention decoding," in \textit{Proc. IEEE/CVF Winter Conference on Applications of Computer Vision (WACV)}, 2023, pp. 6222-6231.


\end{thebibliography}
\end{document}